\documentclass[a4paper,12pt]{article}
\usepackage[a4paper,text={16.8cm,22.4cm}]{geometry}
\usepackage{amsmath,amssymb,bm}
\usepackage{graphicx}
\usepackage{cite}

\allowdisplaybreaks 
\usepackage[implicit=false]{hyperref}

\begin{document}

\begin{flushright}
  MZ-TH/10-31 \\
  September 5, 2010 \\
\end{flushright}
  
\vspace{5ex}

\begin{center}
  \textbf{\Large Updated Predictions for Higgs Production at the Tevatron and the LHC}
  
  \vspace{7ex}
  
  \textsc{Valentin Ahrens$^{a}$,
    Thomas Becher$^{b}$,
    Matthias Neubert$^{a}$,
    and Li Lin Yang$^{a}$}
  
  \vspace{2ex}
    
  \textsl{${}^a$Institut f\"ur Physik (THEP), Johannes Gutenberg-Universit\"at\\
    D-55099 Mainz, Germany\\[0.3cm]
    ${}^b$Institute for Theoretical Physics, University of Bern\\
    CH-3012 Bern, Switzerland}
\end{center}
  
\vspace{3ex}
  
\begin{abstract}
  We present updated predictions for the total cross section for Higgs boson production
  through gluon fusion at hadron colliders. In addition to renormalization-group
  improvement at next-to-next-to-next-to-leading logarithmic accuracy, we incorporate the
  two-loop electroweak corrections, which leads to the most precise predictions at
  present. Numerical results are given for Higgs masses between 115~GeV and 200~GeV at the
  Tevatron with $\sqrt{s}=1.96$~TeV and the LHC with $\sqrt{s}=7$--14~TeV.
\end{abstract}

\vspace{4ex}

The search for the Higgs boson is of the highest priority in the experimental programs at
the Fermilab Tevatron and the CERN LHC. The lower bound for the Higgs mass obtained by the
direct searches at the LEP, $m_H \geq 114.4$~GeV at the 95\% CL, has been around for
several years \cite{Barate:2003sz}. At the beginning of this year, the CDF and D0
Collaborations published a new result which excludes Higgs bosons with a mass around
$2m_W$ \cite{Aaltonen:2010yv}. After a recent update, the Tevatron exclusion now covers
the range $158~{\rm GeV} < m_H < 175~{\rm GeV}$ \cite{:2010ar}. On the other hand, the
electroweak precision measurements favor a relatively light Higgs boson with a mass well
below 200~GeV \cite{Alcaraz:2009jr}. The LHC has started operation recently, and the
standard model Higgs boson, if it exists, should be within reach in the next few years.

At hadron colliders, the most important production channel for the Higgs boson is the
gluon fusion process. Much effort has been devoted to improving the theoretical
predictions for this process, especially since it is well known that the total cross
section suffers from huge QCD corrections \cite{Dawson:1990zj, Djouadi:1991tka,
  Harlander:2002wh, Anastasiou:2002yz, Ravindran:2003um}. In the recent papers
\cite{Ahrens:2008qu, Ahrens:2008nc}, we have pointed out that a large portion of these
corrections comes from enhanced contributions of the form $(C_A\pi\alpha_s)^n$, which
arise in the analytic continuation of the gluon form factor from space-like to time-like
momentum transfer. In those two papers, these large contributions, as well as threshold
enhanced terms, were resummed to all orders in $\alpha_s$ at
next-to-next-to-next-to-leading logarithmic (N$^3$LL) accuracy using renormalization-group
(RG) methods.

It is however necessary to update the numerical predictions presented in
\cite{Ahrens:2008nc}. One reason is that there we have only provided results for the LHC
at $\sqrt{s}=14$~TeV, while it is now clear that the LHC will operate at a lower energy
for two or more years. Another reason is the recent effort to evaluate the electroweak
corrections to this process \cite{Aglietti:2004nj, Degrassi:2004mx, Actis:2008ug,
  Actis:2008ts}. Given that QCD effects are well under control in our result (less than
3\% remaining scale uncertainty and perfect perturbative convergence), these electroweak
corrections, which can be as large as 6\%, are non-negligible and should be included. The
$\mathcal{O}(\alpha)$ electroweak corrections can be split into two parts. The part
involving a light quark loop was computed in \cite{Aglietti:2004nj}. The part involving
the top quark in the loop was first calculated in \cite{Degrassi:2004mx} as an expansion
in $m_H^2/(4m_W^2)$, which is therefore formally valid only for $m_H < 2m_W$. The complete
$\mathcal{O}(\alpha)$ corrections including the exact top quark contributions were later
evaluated in \cite{Actis:2008ug, Actis:2008ts} using numerical methods.

Given the $\mathcal{O}(\alpha)$ corrections, there are still ambiguities in how to combine
them with the QCD corrections. In \cite{Actis:2008ug} two schemes were proposed, which
were called the ``partial factorization'' scheme and the ``complete factorization''
scheme. In the partial factorization scheme the $\mathcal{O}(\alpha)$ corrections are
simply added to the QCD corrected cross section, while in the complete factorization
scheme the $\mathcal{O}(\alpha)$ corrections serve as a prefactor in front of the QCD
corrected cross section, which then generate terms of $\mathcal{O}(\alpha\alpha_s^n)$.
Since the QCD corrections in fixed-order perturbation theory are large, these two schemes
can have non-negligible differences, and it was not known at that time which one is better
without an explicit calculation of the $\mathcal{O}(\alpha\alpha_s)$ contributions. This
task has been undertaken in \cite{Anastasiou:2008tj}, where it was demonstrated that
although the complete factorization does not hold exactly, numerically it gives a good
approximation to the $\mathcal{O}(\alpha\alpha_s)$ terms. We will therefore adopt the
complete factorization approach in our result. The relative contribution of the
electroweak corrections is about 4\% for $m_H \sim 100$~GeV, rises to about 6\% at the
$WW$ threshold, and quickly drops to about $-2\%$ for $m_H \sim 200$~GeV.

The uncertainties in our predictions come from several sources. The uncertainty concerning
unknown higher-order QCD corrections can be estimated from the scale dependence of the
cross section. In our approach there are four scales: $\mu_t$, $\mu_h$, $\mu_s$ and
$\mu_f$, and we estimate the scale uncertainty by varying the scales up and down from
their central values and adding in quadrature the associated variations of the cross
section (see \cite{Ahrens:2008nc} for details). The resulting uncertainty is less than 3\%
for both the Tevatron and the LHC. The uncertainty inherent in the experimental
determinations of the parton distribution functions (PDFs) and the strong coupling
constant $\alpha_s$ can be estimated at 90\% CL using modern PDF sets \cite{Martin:2009bu,
  Lai:2010nw, Demartin:2010er}. It is found to range from 11\% to 15\% for the Tevatron
and is about 8\% for the LHC. On top of these there is a small uncertainty coming from the
use of the heavy top limit in the calculation of QCD corrections, which has been shown to
be a very good approximation for a relatively light Higgs boson at NLO \cite{Spira:1995rr}
and recently also at NNLO \cite{Harlander:2009mq, Pak:2009dg}. We also treat the
perturbative correction to Higgs production via a $b$-quark loop in the heavy quark limit.
This approximation is fairly crude and results in an additional uncertainty of about 1\%
in the total cross section.

\begin{table}[t!]
  \centering

  \begin{tabular}{|c|c|c|c|c|}
    \hline
    $m_H$ [GeV] & Tevatron & LHC (7~TeV) & LHC (10~TeV) & LHC (14~TeV)
    \\ \hline
    115 
    & $1.215^{+0.031+0.141}_{-0.007-0.135}$
    & $18.19^{+0.53+1.46}_{-0.14-1.39}$
    & $33.7^{+1.0+2.6}_{-0.2-2.5}$
    & $57.9^{+1.6+4.4}_{-0.3-4.2}$
    \\ \hline
    120 
    & $1.073^{+0.026+0.126}_{-0.006-0.121}$
    & $16.73^{+0.48+1.34}_{-0.13-1.28}$
    & $31.2^{+0.9+2.4}_{-0.2-2.3}$
    & $54.0^{+1.5+4.1}_{-0.3-3.9}$
    \\ \hline
    125 
    & $0.950^{+0.022+0.113}_{-0.005-0.108}$
    & $15.43^{+0.44+1.23}_{-0.12-1.18}$
    & $29.0^{+0.8+2.2}_{-0.2-2.1}$
    & $50.4^{+1.4+3.8}_{-0.3-3.6}$
    \\ \hline
    130 
    & $0.844^{+0.019+0.102}_{-0.004-0.098}$
    & $14.27^{+0.40+1.14}_{-0.11-1.09}$
    & $27.0^{+0.7+2.1}_{-0.2-2.0}$
    & $47.2^{+1.3+3.5}_{-0.3-3.4}$
    \\ \hline
    135 
    & $0.753^{+0.016+0.093}_{-0.004-0.088}$
    & $13.23^{+0.36+1.06}_{-0.10-1.01}$
    & $25.2^{+0.7+1.9}_{-0.2-1.8}$
    & $44.3^{+1.2+3.3}_{-0.3-3.2}$
    \\ \hline
    140 
    & $0.672^{+0.014+0.084}_{-0.003-0.080}$
    & $12.29^{+0.33+0.98}_{-0.09-0.94}$
    & $23.5^{+0.6+1.8}_{-0.2-1.7}$
    & $41.6^{+1.1+3.1}_{-0.3-3.0}$
    \\ \hline
    145 
    & $0.602^{+0.012+0.076}_{-0.003-0.072}$
    & $11.44^{+0.31+0.91}_{-0.08-0.88}$
    & $22.1^{+0.6+1.7}_{-0.1-1.6}$
    & $39.2^{+1.0+2.9}_{-0.2-2.8}$
    \\ \hline
    150 
    & $0.541^{+0.010+0.070}_{-0.002-0.066}$
    & $10.67^{+0.28+0.85}_{-0.08-0.82}$
    & $20.7^{+0.5+1.6}_{-0.1-1.5}$
    & $37.0^{+1.0+2.7}_{-0.2-2.6}$
    \\ \hline
    155 
    & $0.486^{+0.009+0.064}_{-0.002-0.060}$
    & \phantom{1}$9.95^{+0.26+0.80}_{-0.07-0.77}$
    & $19.4^{+0.5+1.5}_{-0.1-1.4}$
    & $34.9^{+0.9+2.6}_{-0.2-2.5}$
    \\ \hline
    160 
    & $0.433^{+0.008+0.058}_{-0.002-0.054}$
    & \phantom{1}$9.21^{+0.24+0.74}_{-0.07-0.71}$
    & $18.1^{+0.5+1.4}_{-0.1-1.3}$
    & $32.7^{+0.8+2.4}_{-0.2-2.3}$
    \\ \hline
    165 
    & $0.385^{+0.006+0.052}_{-0.002-0.049}$
    & \phantom{1}$8.50^{+0.22+0.68}_{-0.06-0.66}$
    & $16.8^{+0.4+1.3}_{-0.1-1.2}$
    & $30.5^{+0.8+2.2}_{-0.2-2.1}$
    \\ \hline
    170 
    & $0.345^{+0.005+0.047}_{-0.002-0.044}$
    & \phantom{1}$7.89^{+0.20+0.63}_{-0.06-0.61}$
    & $15.7^{+0.4+1.2}_{-0.1-1.1}$
    & $28.6^{+0.7+2.1}_{-0.2-2.0}$
    \\ \hline
    175 
    & $0.310^{+0.005+0.043}_{-0.001-0.040}$
    & \phantom{1}$7.36^{+0.18+0.59}_{-0.05-0.57}$
    & $14.7^{+0.4+1.1}_{-0.1-1.1}$
    & $27.0^{+0.7+1.9}_{-0.2-1.9}$
    \\ \hline
    180 
    & $0.280^{+0.004+0.040}_{-0.001-0.037}$
    & \phantom{1}$6.88^{+0.17+0.56}_{-0.05-0.54}$
    & $13.8^{+0.3+1.0}_{-0.1-1.0}$
    & $25.5^{+0.6+1.8}_{-0.2-1.8}$
    \\ \hline
    185 
    & $0.252^{+0.003+0.036}_{-0.001-0.033}$
    & \phantom{1}$6.42^{+0.15+0.52}_{-0.04-0.50}$
    & $13.0^{+0.3+1.0}_{-0.1-0.9}$
    & $24.0^{+0.6+1.7}_{-0.1-1.7}$
    \\ \hline
    190 
    & $0.228^{+0.003+0.033}_{-0.001-0.031}$
    & \phantom{1}$6.02^{+0.14+0.49}_{-0.04-0.47}$
    & $12.2^{+0.3+0.9}_{-0.1-0.9}$
    & $22.7^{+0.5+1.6}_{-0.1-1.6}$
    \\ \hline
    195 
    & $0.207^{+0.002+0.031}_{-0.001-0.028}$
    & \phantom{1}$5.67^{+0.13+0.46}_{-0.04-0.45}$
    & $11.6^{+0.3+0.9}_{-0.1-0.8}$
    & $21.6^{+0.5+1.6}_{-0.1-1.5}$
    \\ \hline
    200 
    & $0.189^{+0.002+0.028}_{-0.001-0.026}$
    & \phantom{1}$5.35^{+0.12+0.44}_{-0.03-0.42}$
    & $11.0^{+0.3+0.8}_{-0.1-0.8}$
    & $20.6^{+0.5+1.5}_{-0.1-1.4}$
    \\ \hline
  \end{tabular}

  \caption{\label{tbl:mstw} Cross sections (in pb) for different Higgs masses at the
    Tevatron and the LHC, using MSTW2008NNLO PDFs. The first error accounts for scale
    variations, while the second one reflects the combined uncertainty from the PDFs and
    $\alpha_s$.}
\end{table}

\begin{figure}[t!]
  \centering
  \includegraphics[width=0.49\textwidth]{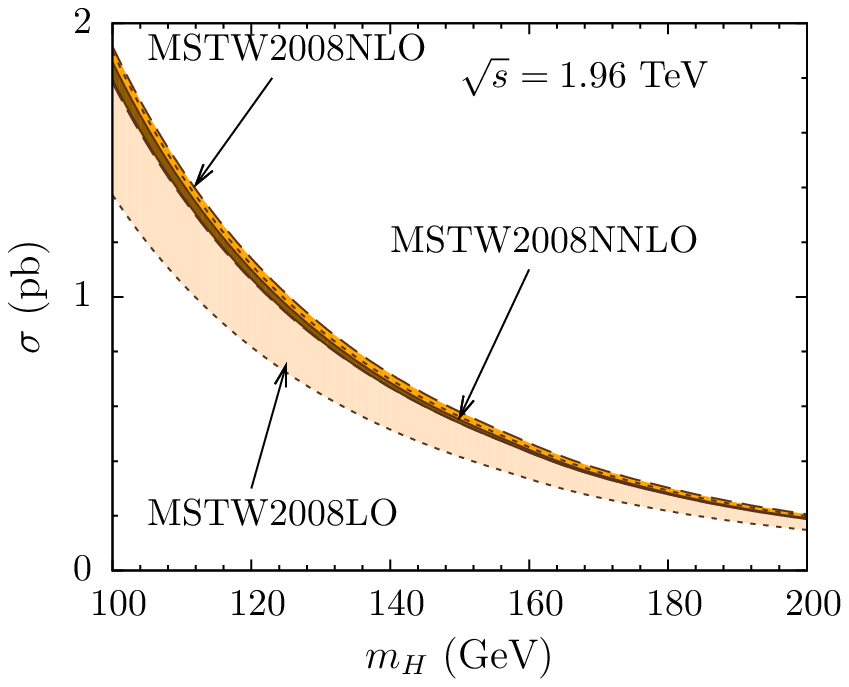}
  \includegraphics[width=0.49\textwidth]{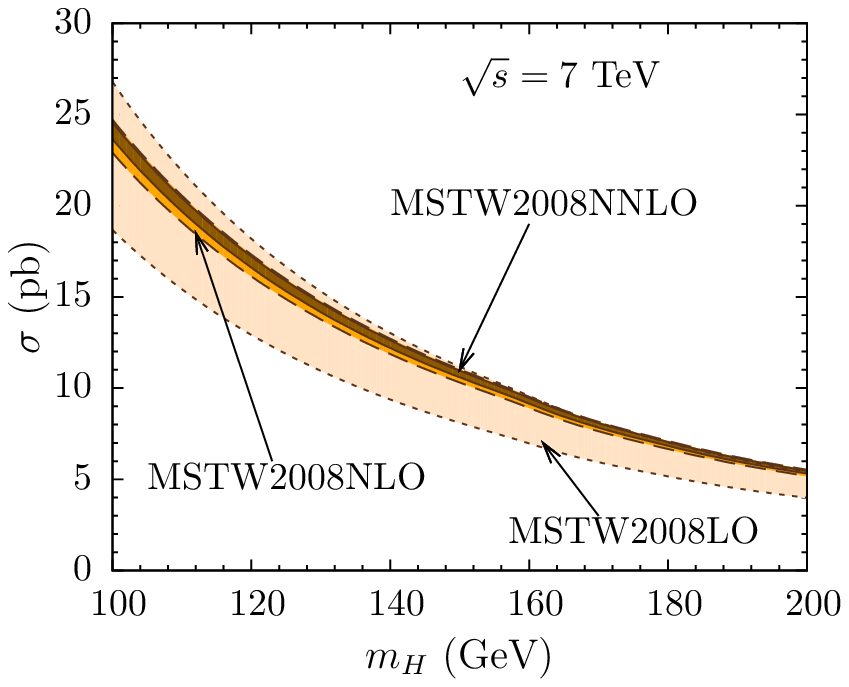}
  \\
  \includegraphics[width=0.49\textwidth]{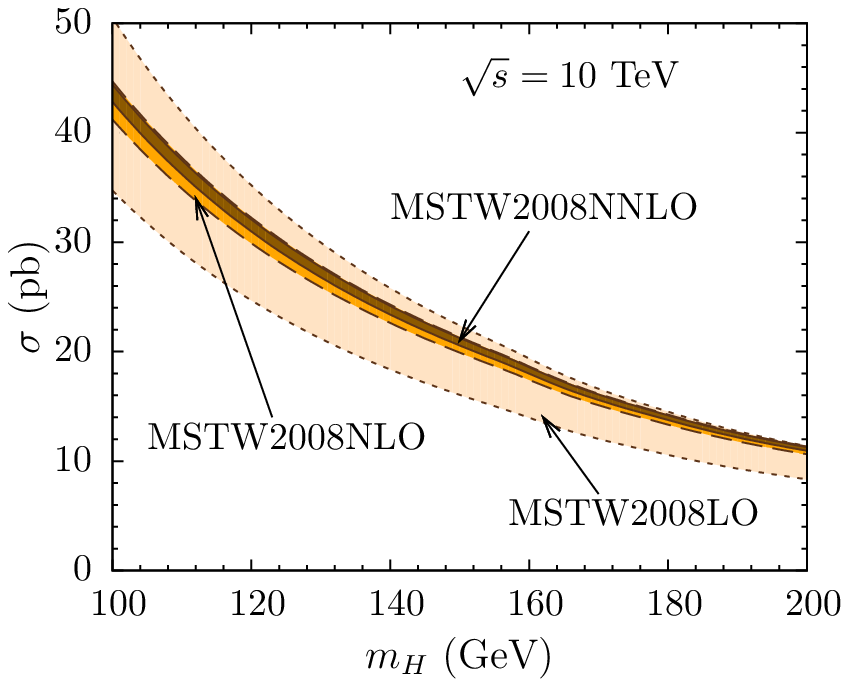}
  \includegraphics[width=0.49\textwidth]{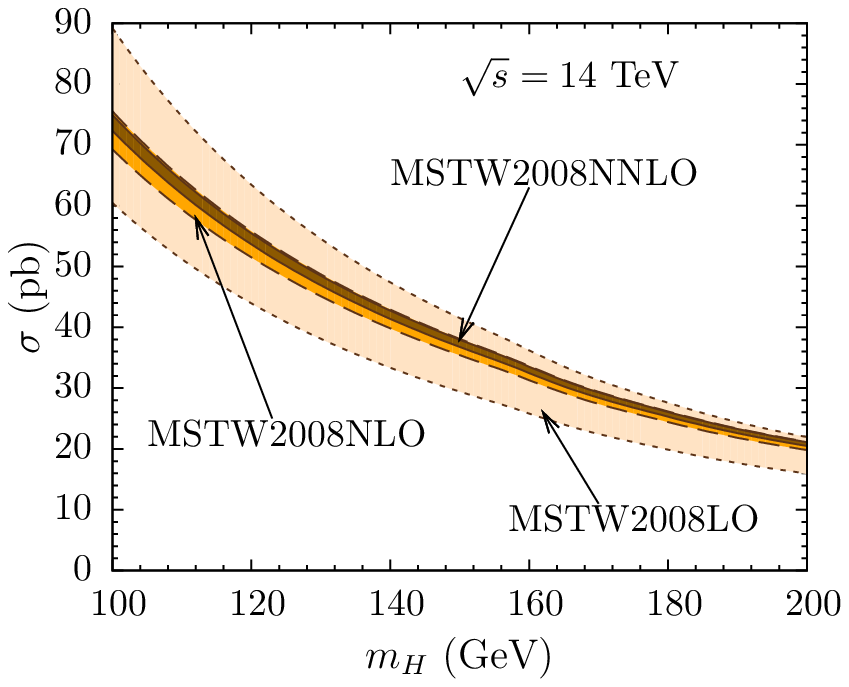}
  \caption{Cross sections at the Tevatron for $\sqrt{s}=1.96$~TeV and the LHC for
    $\sqrt{s}=7$, 10, 14~TeV. Bands indicate scale uncertainties. Light, medium and dark
    bands represent LO (NLL), NLO (NNLL) and NNLO (N$^3$LL) in RG-improved perturbation
    theory, respectively.}
  \label{fig:bands}
\end{figure}

\begin{table}[t!]
  \centering

  \begin{tabular}{|c|c|c|c|c|}
    \hline
    $m_H$ [GeV] & Tevatron & LHC (7~TeV) & LHC (10~TeV) & LHC (14~TeV)
    \\ \hline
    115 
    & $1.215^{+0.031+0.105}_{-0.007-0.095}$
    & $18.34^{+0.54+0.95}_{-0.14-1.00}$
    & $34.1^{+1.0+1.8}_{-0.2-1.9}$
    & $58.8^{+1.7+3.1}_{-0.4-3.5}$
    \\ \hline
    120 
    & $1.073^{+0.026+0.096}_{-0.005-0.087}$
    & $16.86^{+0.49+0.87}_{-0.13-0.91}$
    & $31.5^{+0.9+1.6}_{-0.2-1.8}$
    & $54.7^{+1.6+2.9}_{-0.3-3.2}$
    \\ \hline
    125 
    & $0.950^{+0.022+0.088}_{-0.005-0.079}$
    & $15.54^{+0.45+0.80}_{-0.12-0.83}$
    & $29.3^{+0.8+1.5}_{-0.2-1.6}$
    & $51.1^{+1.4+2.6}_{-0.3-3.0}$
    \\ \hline
    130 
    & $0.845^{+0.019+0.081}_{-0.004-0.072}$
    & $14.36^{+0.41+0.74}_{-0.11-0.76}$
    & $27.2^{+0.8+1.4}_{-0.2-1.5}$
    & $47.8^{+1.3+2.5}_{-0.3-2.7}$
    \\ \hline
    135 
    & $0.753^{+0.016+0.075}_{-0.004-0.067}$
    & $13.31^{+0.37+0.68}_{-0.10-0.70}$
    & $25.4^{+0.7+1.3}_{-0.2-1.4}$
    & $44.8^{+1.2+2.3}_{-0.3-2.5}$
    \\ \hline
    140 
    & $0.673^{+0.014+0.069}_{-0.003-0.061}$
    & $12.35^{+0.34+0.63}_{-0.09-0.65}$
    & $23.7^{+0.7+1.2}_{-0.2-1.3}$
    & $42.1^{+1.1+2.1}_{-0.3-2.3}$
    \\ \hline
    145 
    & $0.604^{+0.012+0.064}_{-0.003-0.057}$
    & $11.50^{+0.31+0.59}_{-0.08-0.60}$
    & $22.2^{+0.6+1.1}_{-0.2-1.2}$
    & $39.7^{+1.1+2.0}_{-0.2-2.2}$
    \\ \hline
    150 
    & $0.542^{+0.010+0.059}_{-0.002-0.052}$
    & $10.71^{+0.29+0.55}_{-0.08-0.56}$
    & $20.9^{+0.6+1.0}_{-0.1-1.1}$
    & $37.4^{+1.0+1.9}_{-0.2-2.0}$
    \\ \hline
    155 
    & $0.487^{+0.009+0.055}_{-0.002-0.049}$
    & \phantom{1}$9.99^{+0.26+0.51}_{-0.07-0.52}$
    & $19.6^{+0.5+1.0}_{-0.1-1.0}$
    & $35.2^{+0.9+1.7}_{-0.2-1.9}$
    \\ \hline
    160 
    & $0.435^{+0.008+0.050}_{-0.002-0.045}$
    & \phantom{1}$9.24^{+0.24+0.48}_{-0.07-0.48}$
    & $18.2^{+0.5+0.9}_{-0.1-0.9}$
    & $33.0^{+0.9+1.6}_{-0.2-1.7}$
    \\ \hline
    165 
    & $0.387^{+0.007+0.046}_{-0.002-0.041}$
    & \phantom{1}$8.52^{+0.22+0.44}_{-0.06-0.44}$
    & $16.9^{+0.4+0.8}_{-0.1-0.9}$
    & $30.7^{+0.8+1.5}_{-0.2-1.6}$
    \\ \hline
    170 
    & $0.347^{+0.006+0.043}_{-0.002-0.038}$
    & \phantom{1}$7.91^{+0.20+0.41}_{-0.05-0.41}$
    & $15.8^{+0.4+0.8}_{-0.1-0.8}$
    & $28.8^{+0.7+1.4}_{-0.2-1.5}$
    \\ \hline
    175 
    & $0.313^{+0.005+0.039}_{-0.001-0.035}$
    & \phantom{1}$7.38^{+0.19+0.38}_{-0.05-0.38}$
    & $14.8^{+0.4+0.7}_{-0.1-0.7}$
    & $27.2^{+0.7+1.3}_{-0.2-1.4}$
    \\ \hline
    180 
    & $0.282^{+0.004+0.037}_{-0.001-0.032}$
    & \phantom{1}$6.89^{+0.17+0.36}_{-0.05-0.36}$
    & $13.9^{+0.3+0.7}_{-0.1-0.7}$
    & $25.7^{+0.6+1.2}_{-0.2-1.3}$
    \\ \hline
    185 
    & $0.254^{+0.004+0.034}_{-0.001-0.030}$
    & \phantom{1}$6.43^{+0.16+0.34}_{-0.04-0.33}$
    & $13.1^{+0.3+0.6}_{-0.1-0.7}$
    & $24.2^{+0.6+1.1}_{-0.1-1.2}$
    \\ \hline
    190 
    & $0.230^{+0.003+0.032}_{-0.001-0.028}$
    & \phantom{1}$6.02^{+0.15+0.32}_{-0.04-0.31}$
    & $12.3^{+0.3+0.6}_{-0.1-0.6}$
    & $22.9^{+0.6+1.1}_{-0.1-1.2}$
    \\ \hline
    195 
    & $0.210^{+0.003+0.030}_{-0.001-0.026}$
    & \phantom{1}$5.67^{+0.14+0.30}_{-0.04-0.30}$
    & $11.6^{+0.3+0.6}_{-0.1-0.6}$
    & $21.8^{+0.5+1.0}_{-0.1-1.1}$
    \\ \hline
    200 
    & $0.191^{+0.002+0.028}_{-0.001-0.024}$
    & \phantom{1}$5.35^{+0.13+0.29}_{-0.03-0.28}$
    & $11.1^{+0.3+0.5}_{-0.1-0.5}$
    & $20.8^{+0.5+1.0}_{-0.1-1.0}$
    \\ \hline
  \end{tabular}

  \caption{\label{tbl:ct10} Cross sections (in pb) for different Higgs masses at the
    Tevatron and the LHC, using CT10 PDFs with $\alpha_s(m_Z)=0.118$.}
\end{table}

\begin{table}[t!]
  \centering
  
  \begin{tabular}{|c|c|c|c|c|}
    \hline
    $m_H$ [GeV] & Tevatron & LHC (7~TeV) & LHC (10~TeV) & LHC (14~TeV)
    \\ \hline
    115 
    & $1.341^{+0.037+0.143}_{-0.018-0.143}$
    & $19.35^{+0.60+1.36}_{-0.29-1.36}$
    & $35.4^{+1.1+2.4}_{-0.5-2.4}$
    & $60.3^{+1.8+3.9}_{-0.7-3.9}$
    \\ \hline
    120 
    & $1.184^{+0.032+0.129}_{-0.016-0.129}$
    & $17.82^{+0.54+1.25}_{-0.29-1.25}$
    & $32.8^{+1.0+2.2}_{-0.5-2.2}$
    & $56.3^{+1.7+3.7}_{-0.7-3.7}$
    \\ \hline
    125 
    & $1.049^{+0.027+0.116}_{-0.014-0.116}$
    & $16.45^{+0.50+1.15}_{-0.28-1.15}$
    & $30.5^{+0.9+2.0}_{-0.5-2.0}$
    & $52.6^{+1.5+3.4}_{-0.8-3.4}$
    \\ \hline
    130 
    & $0.932^{+0.023+0.105}_{-0.013-0.105}$
    & $15.23^{+0.45+1.07}_{-0.28-1.07}$
    & $28.5^{+0.8+1.9}_{-0.5-1.9}$
    & $49.3^{+1.4+3.2}_{-0.8-3.2}$
    \\ \hline
    135 
    & $0.831^{+0.020+0.096}_{-0.011-0.096}$
    & $14.13^{+0.41+0.99}_{-0.27-0.99}$
    & $26.6^{+0.8+1.8}_{-0.5-1.8}$
    & $46.3^{+1.3+3.0}_{-0.8-3.0}$
    \\ \hline
    140 
    & $0.742^{+0.017+0.087}_{-0.010-0.087}$
    & $13.14^{+0.38+0.93}_{-0.26-0.93}$
    & $24.9^{+0.7+1.7}_{-0.5-1.7}$
    & $43.6^{+1.2+2.8}_{-0.8-2.8}$
    \\ \hline
    145 
    & $0.665^{+0.015+0.080}_{-0.009-0.080}$
    & $12.24^{+0.35+0.86}_{-0.25-0.86}$
    & $23.3^{+0.7+1.5}_{-0.5-1.5}$
    & $41.1^{+1.1+2.6}_{-0.8-2.6}$
    \\ \hline
    150 
    & $0.597^{+0.013+0.073}_{-0.008-0.073}$
    & $11.42^{+0.32+0.81}_{-0.24-0.81}$
    & $21.9^{+0.6+1.5}_{-0.4-1.5}$
    & $38.8^{+1.1+2.5}_{-0.7-2.5}$
    \\ \hline
    155 
    & $0.536^{+0.011+0.067}_{-0.007-0.067}$
    & $10.66^{+0.30+0.76}_{-0.23-0.76}$
    & $20.6^{+0.6+1.4}_{-0.4-1.4}$
    & $36.6^{+1.0+2.3}_{-0.7-2.3}$
    \\ \hline
    160 
    & $0.478^{+0.010+0.061}_{-0.006-0.061}$
    & \phantom{1}$9.88^{+0.27+0.70}_{-0.22-0.70}$
    & $19.2^{+0.5+1.3}_{-0.4-1.3}$
    & $34.3^{+0.9+2.2}_{-0.7-2.2}$
    \\ \hline
    165 
    & $0.425^{+0.008+0.055}_{-0.005-0.055}$
    & \phantom{1}$9.11^{+0.25+0.65}_{-0.21-0.65}$
    & $17.8^{+0.5+1.2}_{-0.4-1.2}$
    & $32.0^{+0.9+2.0}_{-0.7-2.0}$
    \\ \hline
    170 
    & $0.380^{+0.007+0.050}_{-0.005-0.050}$
    & \phantom{1}$8.46^{+0.24+0.61}_{-0.19-0.61}$
    & $16.6^{+0.5+1.1}_{-0.4-1.1}$
    & $30.0^{+0.8+1.9}_{-0.6-1.9}$
    \\ \hline
    175 
    & $0.342^{+0.006+0.046}_{-0.004-0.046}$
    & \phantom{1}$7.90^{+0.22+0.57}_{-0.18-0.57}$
    & $15.6^{+0.4+1.0}_{-0.4-1.0}$
    & $28.4^{+0.8+1.8}_{-0.6-1.8}$
    \\ \hline
    180 
    & $0.308^{+0.005+0.042}_{-0.003-0.042}$
    & \phantom{1}$7.38^{+0.20+0.53}_{-0.17-0.53}$
    & $14.7^{+0.4+1.0}_{-0.3-1.0}$
    & $26.8^{+0.7+1.7}_{-0.6-1.7}$
    \\ \hline
    185 
    & $0.277^{+0.005+0.039}_{-0.003-0.039}$
    & \phantom{1}$6.90^{+0.19+0.50}_{-0.16-0.50}$
    & $13.8^{+0.4+0.9}_{-0.3-0.9}$
    & $25.3^{+0.7+1.6}_{-0.6-1.6}$
    \\ \hline
    190 
    & $0.250^{+0.004+0.036}_{-0.002-0.036}$
    & \phantom{1}$6.46^{+0.18+0.47}_{-0.15-0.47}$
    & $13.0^{+0.4+0.9}_{-0.3-0.9}$
    & $23.9^{+0.7+1.5}_{-0.5-1.5}$
    \\ \hline
    195 
    & $0.227^{+0.004+0.033}_{-0.002-0.033}$
    & \phantom{1}$6.08^{+0.17+0.44}_{-0.14-0.44}$
    & $12.3^{+0.4+0.8}_{-0.3-0.8}$
    & $22.8^{+0.6+1.4}_{-0.5-1.4}$
    \\ \hline
    200 
    & $0.207^{+0.003+0.031}_{-0.002-0.031}$
    & \phantom{1}$5.74^{+0.17+0.42}_{-0.13-0.42}$
    & $11.7^{+0.3+0.8}_{-0.3-0.8}$
    & $21.7^{+0.6+1.4}_{-0.5-1.4}$
    \\ \hline
  \end{tabular}

  \caption{\label{tbl:nnpdf} Cross sections (in pb) for different Higgs masses at the
    Tevatron and the LHC, using NNPDF2.0 PDFs with $\alpha_s(m_Z)=0.119$.}
\end{table}


In our numerical evaluation we take the input parameters as \cite{Amsler:2008zzb, :2009ec}
\begin{gather*}
  m_t = 173.1~\text{GeV} \, , \quad \overline{m}_b(\overline{m}_b) = 4.2~\text{GeV} \, ,
  \\
  m_Z = 91.1876~\text{GeV} \, , \quad G_F(m_Z) = 1.16208 \cdot 10^{-5}~\text{GeV}^{-2} \,
  ,
\end{gather*}
and by default use the MSTW2008NNLO PDFs \cite{Martin:2009iq} with $\alpha_s(m_Z) =
0.11707$. The other electroweak parameters are the same as in \cite{Actis:2008ug}. For
comparison, we also show numbers obtained using the CT10 and NNPDF2.0 PDFs
\cite{Lai:2010vv, Ball:2010de} , with the corresponding values of $\alpha_s(m_Z)$. We
note, however, that these are NLO PDFs and therefore less well suited for our calculation.

\begin{figure}[t!]
  \centering
  \includegraphics[width=0.55\textwidth]{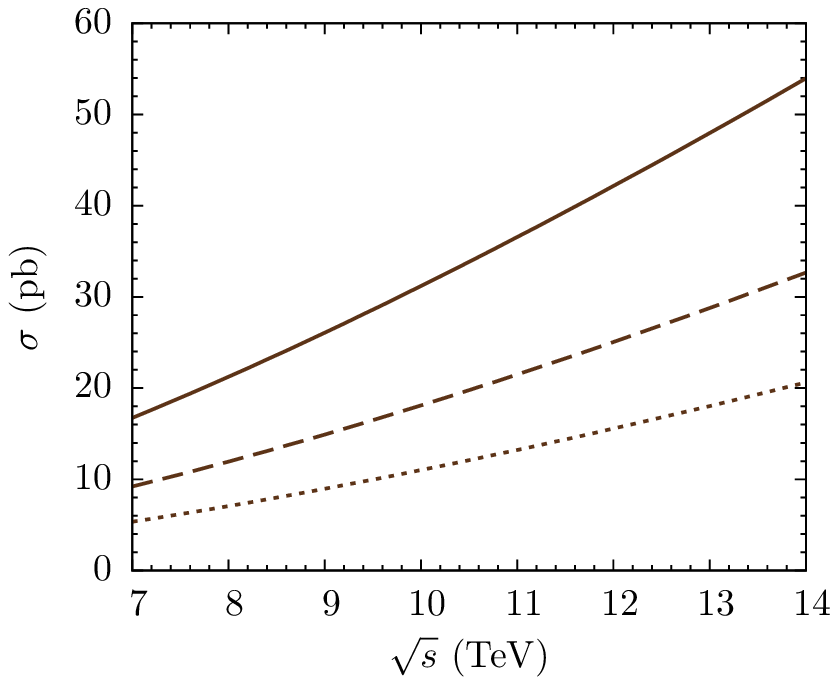}
  \caption{Cross sections as functions of the center-of-mass energy at the LHC for Higgs
    masses being 120~GeV (solid), 160~GeV (dashed) and 200~GeV (dotted).}
  \label{fig:Ecms}
\end{figure}

Our main results are summarized in Table~\ref{tbl:mstw}, where our best predictions for
the cross section at the Tevatron with $\sqrt{s}=1.96$~TeV and the LHC with $\sqrt{s}=7$,
10, 14~TeV using MSTW2008NNLO PDFs are shown. In Figure~\ref{fig:bands}, we show the cross
sections as functions of $m_H$, with bands representing the scale uncertainties. We have
also depicted the LO and NLO RG-improved cross sections in Figure~\ref{fig:bands}, to show
the good perturbative convergence of our result. In Figure~\ref{fig:Ecms}, we plot the
central values of the cross sections at the LHC for $m_H=120$, 160 and 200~GeV as
functions of $\sqrt{s}$. For comparison, in Table~\ref{tbl:ct10} and \ref{tbl:nnpdf} we
also show the cross sections using CT10 and NNPDF2.0 PDFs. They agree with the results in
Table~\ref{tbl:mstw} within errors. To make it simple to update our results in the future,
we include a Fortran program for
download\footnote{\url{http://projects.hepforge.org/rghiggs/}}.

In \cite{deFlorian:2009hc}, the authors have also updated their predictions for Higgs
production via gluon fusion combining soft gluon resummation and two-loop electroweak
corrections. Our results differ in several important aspects from theirs:
\begin{itemize}
\item We work at N$^3$LL accuracy rather than NNLL.
\item We resum the enhanced contributions arising from the analytic continuation of the
  gluon form factor. This has been demonstrated to greatly improve the perturbative
  convergence.
\item We work directly in momentum space rather than in Mellin moment space, which avoids
  the Landau pole ambiguity.
\end{itemize}
Therefore, we believe that our results are the most precise predictions for the total
Higgs production cross sections to date. With the higher-order perturbative corrections
under control, the main uncertainties now arise from the experimental determinations of
the PDFs and $\alpha_s$.

\section*{Acknowledgments}

We are grateful to Alessandro Vicini, Roberto Bonciani and Juan Rojo for useful
discussions. We thank Stefano Actis for pointing us to the grid files for the electroweak
corrections.

\end{document}